\documentclass[apj]{emulateapj}			%

\usepackage{epsfig}
\usepackage{natbib}                
\usepackage{verbatim}              
\usepackage{graphicx}              
\usepackage{amsmath,amsthm,amsfonts,amsopn,amssymb} 
\usepackage{afterpage}
\usepackage{rotating}					

\shorttitle{The GJ~436 System: Astrophysical Parameters of M-Dwarf and Transiting Neptune}

\shortauthors{von Braun \& Boyajian et al.}
\begin{document}


\title{The GJ~436 System: Directly Determined Astrophysical Parameters of an M-Dwarf and Implications for the Transiting Hot Neptune} 

\author{Kaspar von Braun\altaffilmark{1,12},
Tabetha S. Boyajian\altaffilmark{2,3}, 
Stephen R. Kane\altaffilmark{1},
Leslie Hebb\altaffilmark{9},
Gerard T. van Belle\altaffilmark{4},  
Chris Farrington\altaffilmark{6}, 
David R. Ciardi\altaffilmark{1},  
Heather A. Knutson\altaffilmark{11},
Theo A. ten Brummelaar\altaffilmark{6}, 
Mercedes L\'{o}pez-Morales\altaffilmark{5,8},
Harold A. McAlister\altaffilmark{2}, 
Gail Schaefer\altaffilmark{6},
Stephen Ridgway\altaffilmark{7}, 
Andrew Collier Cameron\altaffilmark{10},
P. J. Goldfinger\altaffilmark{6}, 
Nils H. Turner\altaffilmark{6},
Laszlo Sturmann\altaffilmark{6}, 
and Judit Sturmann\altaffilmark{6} 
}

\altaffiltext{1}{NASA Exoplanet Science Institute, California Institute of Technology, MC 100-22, Pasadena, CA 91125}
\altaffiltext{2}{Center for High Angular Resolution Astronomy and Department of Physics and Astronomy, Georgia State University, P. O. Box 4106, Atlanta, GA 30302-4106} 
\altaffiltext{3}{Hubble Fellow} 
\altaffiltext{4}{Lowell Observatory, 1400 W. Mars Hill Road, Flagstaff, AZ 86001}
\altaffiltext{5}{Institut de Ci\`{e}ncies de L'Espai (CSIC-IEEC), Campus UAB, Facultat Ci\`{e}ncies, Torre C5 parell 2, 08193 Bellaterra, Barcelona, Spain}
\altaffiltext{6}{The CHARA Array, Mount Wilson Observatory, Mount Wilson, CA 91023}
\altaffiltext{7}{National Optical Astronomy Observatory, P.O. Box 26732, Tucson, AZ 85726-6732}
\altaffiltext{8}{Department of Terrestrial Magnetism, Carnegie Institution of Washington, 5241 Broad Branch Road, NW, Washington, DC 20015}
\altaffiltext{9}{Vanderbilt University, Nashville, TN}
\altaffiltext{10}{School of Physics \& Astronomy, University of St. Andrews, North Haugh, Fife, KY16 9SS, UK}
\altaffiltext{11}{Dept of Geological and Planetary Sciences, California Institute of Technology, MC 150-21, Pasadena, CA 91125}
\altaffiltext{12}{kaspar@caltech.edu}



\begin{abstract}

The late-type dwarf GJ~436 is known to host a transiting Neptune-mass planet in a 2.6-day orbit. We present results of our interferometric measurements to directly determine the stellar diameter ($R_{\star} = 0.455 \pm 0.018 R_{\odot}$) and effective temperature ($T_{\rm EFF} = 3416 \pm 54$ K). We combine our stellar parameters with literature time-series data, which allows us to calculate physical and orbital system parameters, including GJ~436's stellar mass ($M_{\star} = 0.507^{+ 0.071 }_{- 0.062 } M_{\odot}$) and density ($\rho_* = 5.37^{+ 0.30 }_{- 0.27 } \rho_\odot$), planetary radius ($R_{p} = 0.369^{+ 0.015 }_{- 0.015 } R_{Jupiter}$), planetary mass ($M_{p} = 0.078^{+ 0.007 }_{- 0.008 } M_{Jupiter}$), implying a mean planetary density of $\rho_{p} = 1.55^{+ 0.12 }_{- 0.10 } \rho_{Jupiter}$. These values are generally in good agreement with previous literature estimates based on assumed stellar mass and photometric light curve fitting. Finally, we examine the expected phase curves of the hot Neptune GJ~436b, based on various assumptions concerning the efficiency of energy redistribution in the planetary atmosphere, and find that it could be constrained with {\it Spitzer} monitoring observations.

\end{abstract}



\section{Introduction}\label{sec:introduction}


GJ~436 is an M3 dwarf \citep{kir91,haw96} that is known to host a Neptune-sized exoplanet in a 2.64-day orbit \citep{cac09,bal10b,sou10}. The planet was originally discovered by the radial velocity method  \citep{but04}, and subsequent photometric studies used transit photometry to determine the planet's radius and density: see \citet{demory07}, \citet{deming07}, \citet{gil07b}, \citet{gil07a}, \citet{bea08}, \citet{pon09}, \citet{fig09}, and references therein. One implicit assumption in the calculation of planetary radius and density is the knowledge of the stellar radius calculated from, for instance, stellar models. Particularly in the M dwarf mass regime, however, there is a well-documented discrepancy between model radii and the ones that can be directly determined \citep{boy10}. For spectral types around M3V, the offset between model radii and directly measured counterparts is on the order of 10\% \citep[][and references therein]{tor07,mercedes2007,lop07,von08,boy10}; but see also \citet{dem09}.


The advent of long-baseline interferometry at wavelengths in the near-infrared or optical range has made it possible to circumvent assumptions of stellar radius by enabling direct measurements of stellar radius and other astrophysical properties for nearby, bright stars \citep[e.g., ][and references therein]{bai08,bai08a,bai09,bai10,van09,von11a,von11b,boy11}. Currently the only stars known to host a transiting exoplanet with directly determined radii are HD~189733 \citep{bai07} and 55~Cancri \citep{von11c}. In this paper, we use interferometric observations to obtain GJ~436's astrophysical parameters and provide a physical characterization of the system. We describe our observations in \S \ref{sec:observations} and discuss directly determined and derived stellar and planetary astrophysical properties in \S \ref{sec:properties} and \S \ref{sec:mcmc}, respectively. 
We summarize and conclude in \S \ref{sec:conclusion}.



\section{Interferometric Observations}\label{sec:observations}


\begin{deluxetable*}{rccc}
\tabletypesize{\tiny}
\tablewidth{0pt}
\tablecaption{Observation Log\label{tab:observations}}
\tablehead{
\colhead{\textbf{UT}} &
\colhead{\textbf{ }} &
\colhead{\textbf{\# of}} &
\colhead{\textbf{Calibrator}} \\
\colhead{\textbf{Date}} &
\colhead{\textbf{Baseline}} &
\colhead{\textbf{Brackets}} &
\colhead{\textbf{HD}}
}

\startdata

2011/01/21 & S1/E1 & 6 & HD~95804, HD~102555 \\
2011/01/22 & S1/W1 & 3 & HD~95804 \\
2011/01/24 & S1/E1 & 7 & HD~95804, HD~102555, HD~104349 \\
2011/01/25 & S1/E1 & 3 & HD~102555, HD~104349 \\
2011/01/25 & S1/W1 & 2 & HD~104349 \\

\enddata
\end{deluxetable*}


Our observational methods and strategy are described in detail in \citet{von11a}. We repeat aspects specific to the GJ~436 observations below.

GJ~436 was observed over four nights in January 2011 using the Georgia State University Center for High Angular Resolution Astronomy (CHARA) Array \citep{ten05}, a long baseline interferometer located at Mount Wilson Observatory in Southern California. Two of CHARA's longest baselines, S1E1 (330~m) and S1W1 (278~m), were used to collect the observations in $H$-band ($\lambda_{central} = 1.67$~$\mu$m) with the CHARA Classic beam combiner \citep{stu03,ten05} in single-baseline mode. Table \ref{tab:observations} lists the observations, each of which contains around 2.5 minutes of integration and 1.5 minutes of telescope slewing per object (target and calibrator -- see below). During our observations, GJ~436 was between 60 and 80 degrees elevation.

The interferometric observations employed the common technique of taking bracketed sequences of the object and calibrator stars to remove the influence of atmospheric and instrumental systematics.  We rotated between four calibrators over the observation period to minimize any systematic effects in measuring the diameter of GJ~436, whose faintness and small angular size make it a non-trivial system to observe with CHARA.  The calibrator stars used in our observations were: HD~95804 (spectral type A5; $\theta_{EST} = 0.211\pm0.009$~milliarcseconds [mas]), HD~102555 (F2; $\theta_{EST} = 0.220\pm0.007$~mas), HD~103676 (F2; $\theta_{EST} = 0.267\pm0.009$~mas), HD~104349 (K1~III; $\theta_{EST} = 0.263\pm0.008$~mas)\footnote{$\theta_{EST}$ corresponds the estimated angular diameter of the calibrator stars based on spectral energy distribution fitting.}. As in \citet{von11a,von11c}, calibrator stars were chosen to be near-point-like sources of similar brightness as GJ~436 and located at small angular distances from it. 

\begin{figure}										
\centering
\epsfig{file=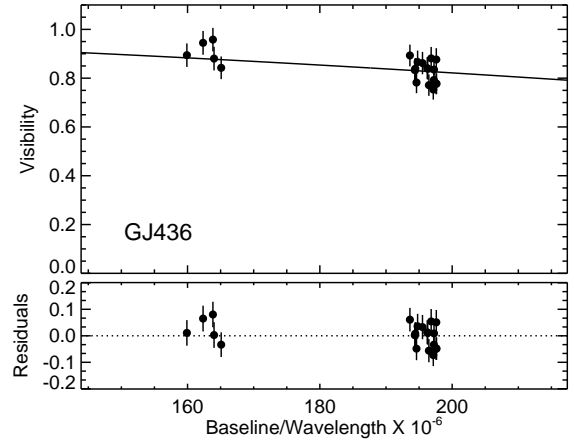,width=\columnwidth,clip=} \\
\caption{Calibrated visibility observations along with the limb-darkened angular diameter fit for GJ~436. Note the two different baseline lengths. For details, see \S \ref{sec:observations}.}
\label{fig:diameters}
\end{figure}

The uniform disk and limb-darkened angular diameters ($\theta_{\rm UD}$ and $\theta_{\rm LD}$, respectively; see Table \ref{tab:properties}) are found by fitting the calibrated visibility measurements (Fig. \ref{fig:diameters}) to the respective functions for each relation.  These functions may be described as $n^{th}$-order Bessel functions that are dependent on the angular diameter of the star, the projected distance between the two telescopes and the wavelength of observation (see equations 2 and 4 of \citealt{han74}). Visibility is the normalized amplitude of the correlation of the light from two telescopes. It is a unitless number ranging from 0 to 1, where 0 implies no correlation, and 1 implies perfect correlation. An unresolved source would have a perfect correlation of 1.0 independent of the distance between the telescopes (baseline). A resolved object will show a decrease in visibility with increasing baseline length. The shape of the visibility versus baseline is a function of the topology of the observed object (the Fourier Transform of the object's shape). For a uniform disk this function is a Bessel function, and for this paper, we use a simple model of a limb-darkened variation of a uniform disk. The visibility of any source is reduced by a non-perfect interferometer, and the point-like calibrators are needed to calibrate out the loss of coherence caused by instrumental effects. We use the linear limb-darkening coefficient $\mu_{H} = 0.3688$ from the PHOENIX models in \citet{cla00} for stellar $T_{\rm EFF}$ = 3400~K and $\log g$ = 5.0 to convert from $\theta_{\rm UD}$ to $\theta_{\rm LD}$. The uncertainties in the adopted limb darkening coefficient amount to 0.2\% when modifying the adopted gravity by 0.5~dex or the adopted $T_{\rm EFF}$ by 200K, well within the errors of our diameter estimate.  Finally, we calculate the effect of baseline smearing due to the finite diameters of the telescope, and we find that its magnitude is around two orders of magnitude below our error estimates.

Our interferometric measurements yield the following values for GJ~436's angular diameters: $\theta_{\rm UD} = 0.405 \pm 0.013$~mas and $\theta_{\rm LD} = 0.417 \pm 0.013$~mas (Table~\ref{tab:properties}).



\section{Directly Determined Parameters}\label{sec:properties}



In this Section, we present our direct measurements of GJ~436's stellar diameter, $T_{\rm EFF}$, and luminosity, based on our interferometric measurements and spectral energy distribution (SED) fitting to literature photometry. Literature stellar and planetary astrophysical parameters for the GJ~436 system, based on spectroscopic analysis, calibrated photometric relations, and time-series photometry data, can be found in the works of, e. g., \citet{gil07b}, \citet{deming07}, \citet{man07}, \citet{tor07}, \citet{bea08}, \citet{cou08},  \citet{bal10a}, \citet{ste10}, \citet{bea11}, \citet{knu11} and \citet{sou08a,sou08b,sou10}. 


\subsection{Stellar Diameter from Interferometry}\label{sec:radius}


\begin{deluxetable*}{rcc}

\tablecaption{Directly Determined and Literature Stellar Properties of GJ 436 \label{tab:properties}} 
\tablewidth{0pc}
\tablehead{
\colhead{Parameter} &
\colhead{Value} &
\colhead{Reference}	
}
\startdata
Spectral Type \dotfill	&	M3V	&	\citet{kir91,haw96} 	\\ 
Parallax (mas) \dotfill				&	$98.61 \pm 2.33$\phn	&	\citet{van07}	\\
$V-K$ \dotfill			&	$4.513$	&	\citet{bes00, cut03}	\\
$\theta_{\rm UD}$ (mas) \dotfill		&	$0.405 \pm 0.013$	&	this work (\S \ref{sec:observations})	\\
$\theta_{\rm LD}$ (mas)	\dotfill	&	$0.417 \pm 0.013$	&	this work (\S \ref{sec:observations})\\
Radius $R_*$ ($R_{\rm \odot}$) \dotfill	&	$0.455 \pm 0.018$	&	this work (\S \ref{sec:radius})\\
Luminosity ($L_{\rm \odot}$) \dotfill	& $0.0253 \pm 0.0012$	&	this work (\S \ref{sec:teff})\\
$T_{\rm EFF}$ (K)	\dotfill			&	$3416 \pm 54$\phn\phn		&	this work (\S \ref{sec:teff})\\
\enddata
\tablecomments{For details, see \S \ref{sec:properties}.}
\end{deluxetable*}


Based on GJ~436's limb-darkened angular diameter $0.417 \pm 0.013$ mas (\S \ref{sec:observations}) and trigonometric parallax from \citet{van07}, we obtain a directly determined physical radius of $R_{\star} = 0.455 \pm 0.018 R_{\rm \odot}$ (see Table \ref{tab:properties}). 

When comparing this value to the radius predicted by the \citet{bar98} stellar models\footnote{For this simple calculation, we assume a 5 Gyr age, solar metallicity, and a 0.44 $M_{\rm \odot}$ mass for GJ~436.}, we reproduce the radius discrepancy for M dwarfs mentioned above and shown in \citet{boy10}: our directly determined radius ($0.455 R_{\rm \odot}$) exceeds the theoretical one ($0.409 R_{\rm \odot}$) by 11\%.  This aspect is further discussed in \citet{tor07}, where the radius of GJ~436 is found to be overly inflated for its mass. 

Otherwise indirectly calculating the stellar radius of GJ~436 requires assumption or knowledge of the stellar mass, published values of which in \citet{man07} and \citet{tor07} are consistent with each other. \citet{man07} estimate a mass of $0.44 \pm 0.04 M_{\rm \odot}$, derived from the empirically driven mass-luminosity relation in \citet{del00}. Following methods outlined in \citet{sea03} and \citet{soz07}, \citet{tor07} calculates GJ~436's mass ($0.452 ^{+0.014}_{-0.012} M_{\rm \odot}$) and radius ($0.464^{+0.011}_{-0.009} R_{\rm \odot}$) by simultaneously applying observational constraints along with adjusting a correction factor to resolve the differences seen when comparing results to models.  

Based on light curve analysis of the transiting planet and the stellar mass values from \citet{man07} or \citet{tor07}, literature values of the stellar radius of GJ~436 reported in \citet[$0.440 \pm 0.04 R_{\rm \odot}$]{gil07a}, \citet[$0.463 \pm 0.022 R_{\rm \odot}$]{gil07b}, \citet[$0.47 \pm 0.02 R_{\rm \odot}$]{deming07}, \citet[$0.45 \pm 0.02 R_{\rm \odot}$]{shp09}, \citet[$0.437 \pm 0.016 R_{\rm \odot}$]{bal10a}, \citet[$0.454 \pm 0.029 R_{\rm \odot}$]{sou10}\footnote{See also: http://www.astro.keele.ac.uk/jkt/tepcat/homogeneous-par-err.html.}, and \citet[$0.437 \pm 0.005 R_{\rm \odot}$]{knu11}, produce stellar radius values that are agreement with our result. Only the radius determined in \citet[$0.505^{+0.029}_{-0.020} R_{\rm \odot}$]{bea08}, based on light curve analysis and the parameters in \citet{man07}, is slightly discrepant ($\simeq 2\sigma$). 
We refer the reader to the excellent \citet{tor07} paper for more background and details on the assumptions and techniques employed by the studies mentioned above.   
The agreements between calculated values and our interferometric radius clearly illustrate the usefulness of exoplanet light curve analysis for the determination of stellar parameters. 


The question can be asked whether an interferometric measurement obtained during planetary transit or during the presence of star spots could yield a radius estimate that is thus artificially reduced. As seen in \citet{van08a}, the effect on observed visibility amplitude of a transiting planet is expected to be very small ($\delta V < 0.005$ predicted for HD189733; GJ~436 with its similar size will have similar magnitude effects).  While a closure phase signal may be detectable in the near future ($\delta CP \simeq 0.2\deg$), visibility variations in a single data point due to a transiting planet will impact current diameter measurements at the $\delta \theta \simeq 0.6$\% level, buried in the measurement noise of the result based on all of our visibility measurements. Furthermore, planetary transits feature a contrast between the stellar surface and the obscuring or darkening feature that is significantly higher than for star spots. Thus, for spots with lower contrast ratios, expected visibility variations are even smaller.


\subsection{Stellar Effective Temperature and Luminosity from SED Fitting}\label{sec:teff}

\begin{figure}										%
\centering
\epsfig{file=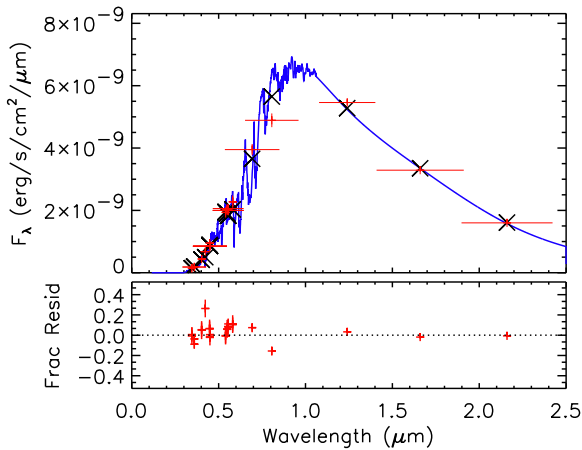,width=\columnwidth,clip=} \\
\caption{SED fit for GJ~436. The (blue) spectrum is a M3V spectral template \citep{pic98}. The (red) crosses indicate photometry values from the literature. ``Error bars'' in x-direction represent bandwidths of the filters used. The (black) X-shaped symbols show the flux value of the spectral template integrated over the filter transmission. The lower panel shows the residuals around the fit in fractional flux units of photometric uncertainty. For details, see \S \ref{sec:teff}.}
\label{fig:sed}
\end{figure}

We produce a fit of the stellar SED based on the spectral templates of \citet{pic98} to literature photometry published in \citet{gol72}, \citet{ruf76}, \citet{mer86}, \citet{sta86}, \citet{doi91}, \citet{leg92}, \citet{wei93}, \citet{wei96}, and \citet{cut03}. In our fit, interstellar extinction is a free parameter and calculated to be $A_V = 0.000 \pm 0.014$ mag, consistent with expectations for a nearby star. The value for the distance to GJ~436 is adopted from \citet{van07}. The SED fit for GJ~436, along with its residuals, is shown in Fig. \ref{fig:sed}.

From the SED fit, we calculate the value of GJ~436's stellar bolometric flux to be $F_{\rm BOL} = (7.881 \pm 0.0497)\times10^{-9}$~erg cm$^{-2}$ s$^{-1}$, and consequently, its luminosity of $L = 0.0253 \pm 0.0012 L_{\odot}$. Combination with the rewritten version of the Stefan-Boltzmann Law

\begin{equation} \label{eq:temperature}
T_{\rm EFF} ({\rm K}) = 2341 (F_{\rm BOL}/\theta_{\rm LD}^2)^{\frac{1}{4}},
\end{equation}

\noindent
where $F_{\rm BOL}$ is in units of $10^{-8}$~erg cm$^{-2}$ s$^{-1}$ and $\theta_{\rm LD}$ is in units of mas, produces GJ~436's effective temperature to be $T_{\rm EFF} = 3416 \pm 54$ K (Table \ref{tab:properties}). Due to the grazing nature of the transit in the GJ~436 system, 
the radius calculated for the planet is more dependent upon the limb-darkening models than the equivalent for central transits. Knowing the stellar effective temperature to a higher precision than before thus provides particularly important constraints for this system.


Using the approach in \citet{von11a,von11c}, we calculate the system's habitable zone to be located at 0.16 -- 0.31 astronomical units (AU) from GJ~436, clearly beyond the orbit of GJ~436b ($a \simeq 0.03$ AU; \S \ref{sec:phasecurve} \& Table \ref{tab:params}.).


\section{Derived Parameters and Results}\label{sec:mcmc}


 In this Section, we use our directly measured stellar parameters (\S \ref{sec:properties} and Table \ref{tab:properties}) and combine them with a global analysis of literature time-series photometry and radial velocity (RV) data to obtain a characterization of the system as a whole, including stellar and planetary physical and orbital parameters. Rather than being required to assume a stellar mass to calculate stellar radius, we are in the position to take our measured value for $R_{\star}$ and calculate a stellar mass.  In addition, we simulate the thermal phase curve of GJ~436b based on our calculated system parameters.

\subsection{MCMC Analysis}\label{sec:analysis}


In a transiting exoplanet system, the shape of the light curve depends in part on the mean stellar density \citep[e.g.,][]{sea03,tin11}. \citet{soz07} illustrated this now widely-used method as an alternative to spectroscopically determined surface gravity, $\log{g}$, to derive more precise system parameters for the host star and its transiting planet. This technique thus enables comparing the mean stellar density and any additionally spectroscopically determined temperature and metallicity to stellar evolution models \citep[e.g.,][]{heb10,eno10}. 


We follow the technique described in \citet{cameron2007} and combine publicly available\footnote{NASA Exoplanet Archive \\ (http://exoplanetarchive.ipac.caltech.edu).} RV and light curve data on GJ~436, along with data from \citet[][Pont, private communication]{pon09}, in a global Markov Chain Monte Carlo (MCMC) analysis, thereby setting our interferometric stellar radius, effective temperature, and associated uncertainties to be fixed. Thus, we are able to make an independent measurement of the mass of this late-type star and realistic associated uncertainty. 

We use 4th order limb darkening coefficients from \citet{cla11} for the IRAC4 and Johnson-Cousins $V$ and $R$-band, for $\log g = 4.5$ and [Fe/H]=0.0 \citep{roj10}. We interpolate the values at GJ~436's effective temperature (\S \ref{sec:teff}) and surface gravity (determined iteratively in the MCMC program) and perform several independent MCMC runs to explore the effect of the adopted limb darkening coefficients on the derived stellar and planetary parameters. We test values generated from both the ATLAS and Phoenix model atmospheres and try two different values for the microturbulence (1.0 and 2.0 km/s).  We find that the different limb darkening values do not introduce any significant variations to the derived parameters. Our results are based on ATLAS model atmospheres, $T_{\rm EFF}$ = 3416 K, $\log g$  interpolated at 4.8, and microturbulence velocity of 1.0 km/s. The limb darkening coefficients we use in our MCMC analysis are given in Table \ref{tab:coeffs}.

\begin{figure*}
 \begin{center}
   \begin{tabular}{cc}
     \includegraphics[angle=0,width=8.2cm]{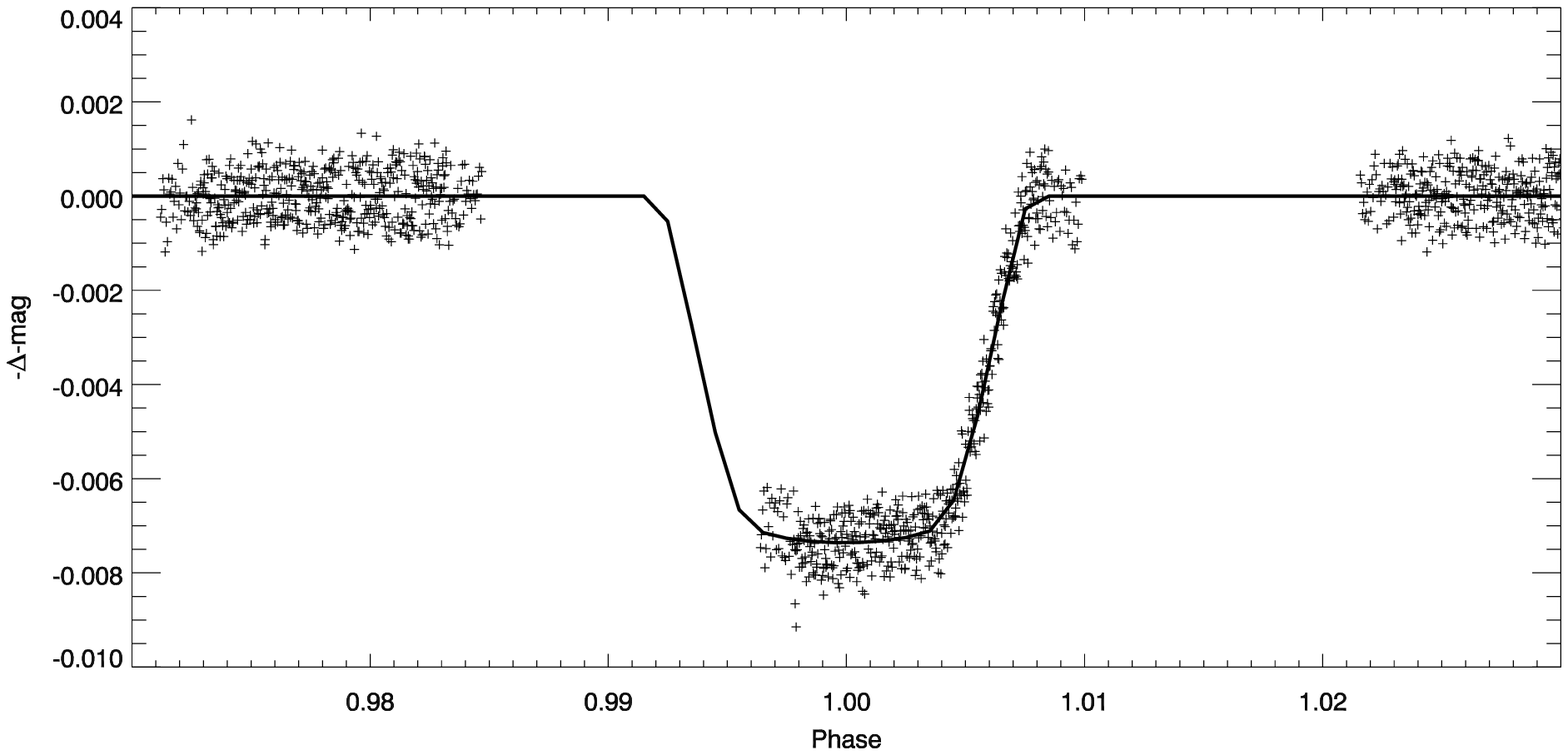} &
     \includegraphics[angle=0,width=8.2cm]{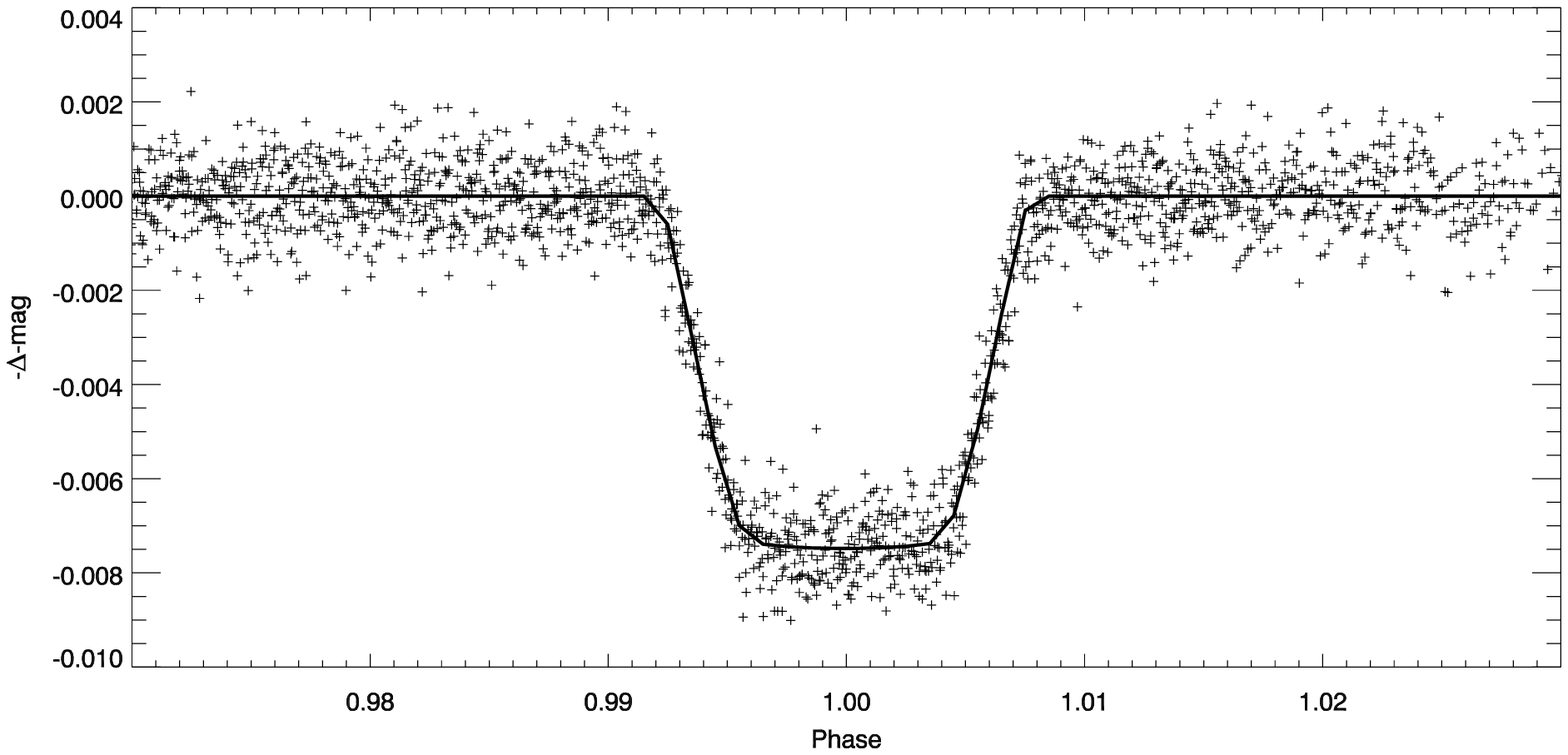} \\     
     \includegraphics[angle=0,width=8.2cm]{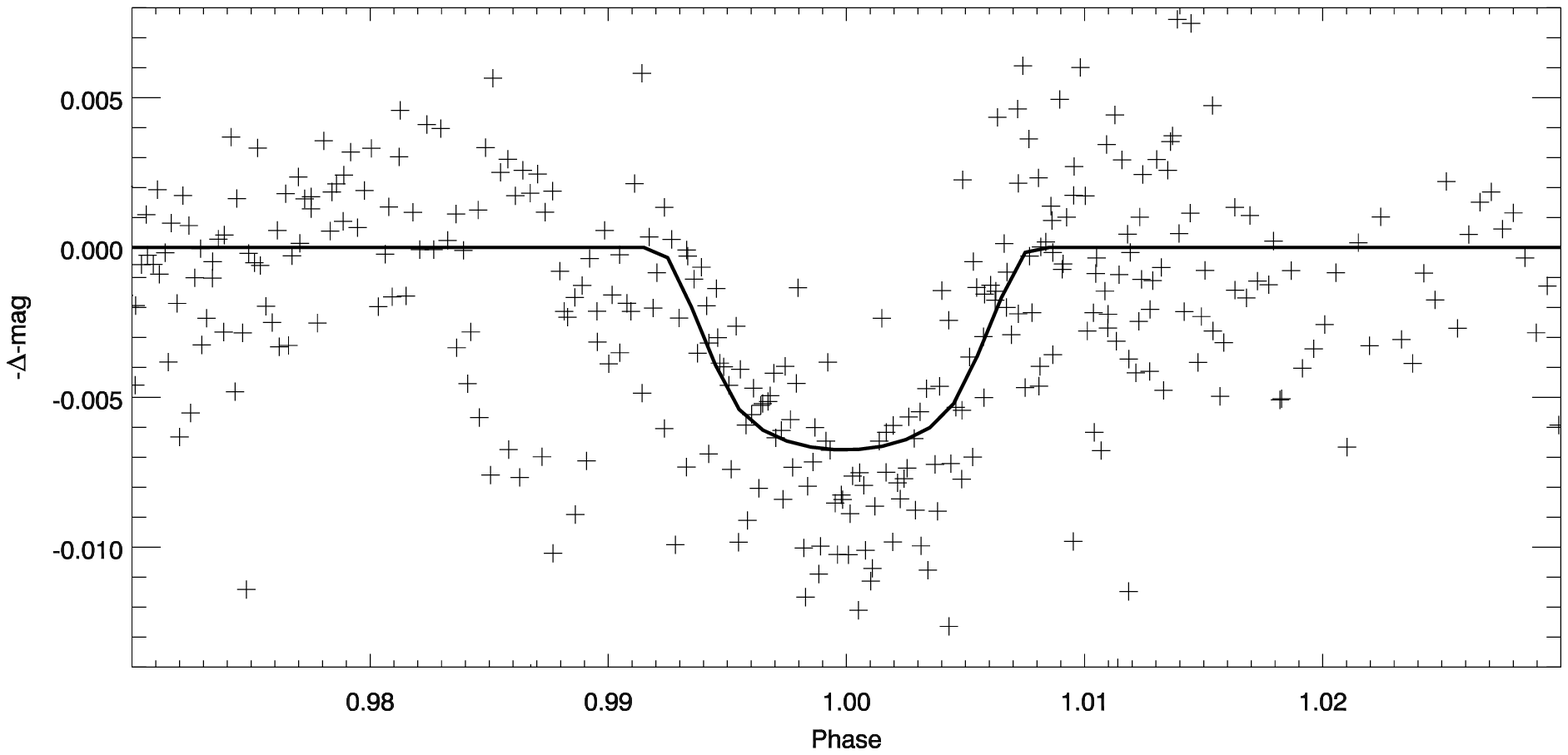} &
\includegraphics[angle=0,width=8.2cm]{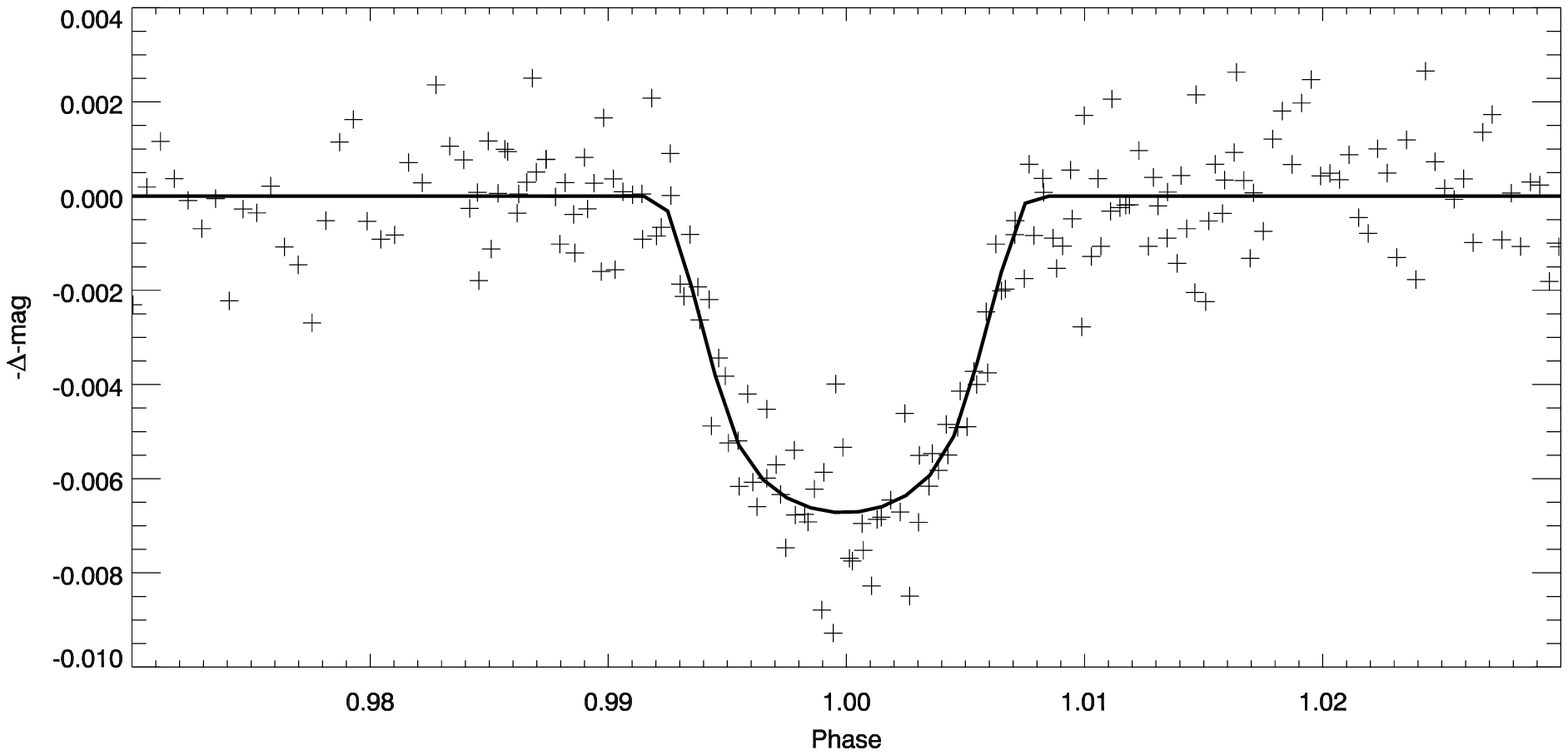} \\

   \end{tabular}
 \end{center}
 \caption{Literature light curves, phase folded with ephemeris information in Table~\ref{tab:params}. The respective model fit generated by our MCMC Analysis is superimposed in every panel. Top left panel: {\it HST} NICMOS data from \citet{pon09}; top right: IRAC-4 primary transit light curve data from \citet{deming07} and \citet{knu11}; bottom left: ground-based $R$ data from \citet{gil07b,gil07a}; bottom right: ground-based $V$ data from \citet{gil07b,gil07a}. For details, see \S \ref{sec:mcmc}.}
\label{fig:lcs_1}
\end{figure*}

\begin{figure}
 \begin{center}
     \includegraphics[angle=0,width=\columnwidth]{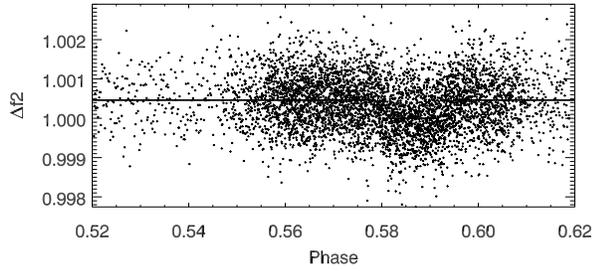} \\
 \end{center}
 \caption{
     IRAC-4 (8-micron) secondary
     eclipse light curve data from \citet{demory07} and \citet{knu11}, phase folded using the ephemeris information in Table~\ref{tab:params}, with our planet model overplotted.
     For details, see \S \ref{sec:mcmc}.}
\label{fig:lcs_2}
\end{figure}

\begin{figure}										%
\centering
\epsfig{file=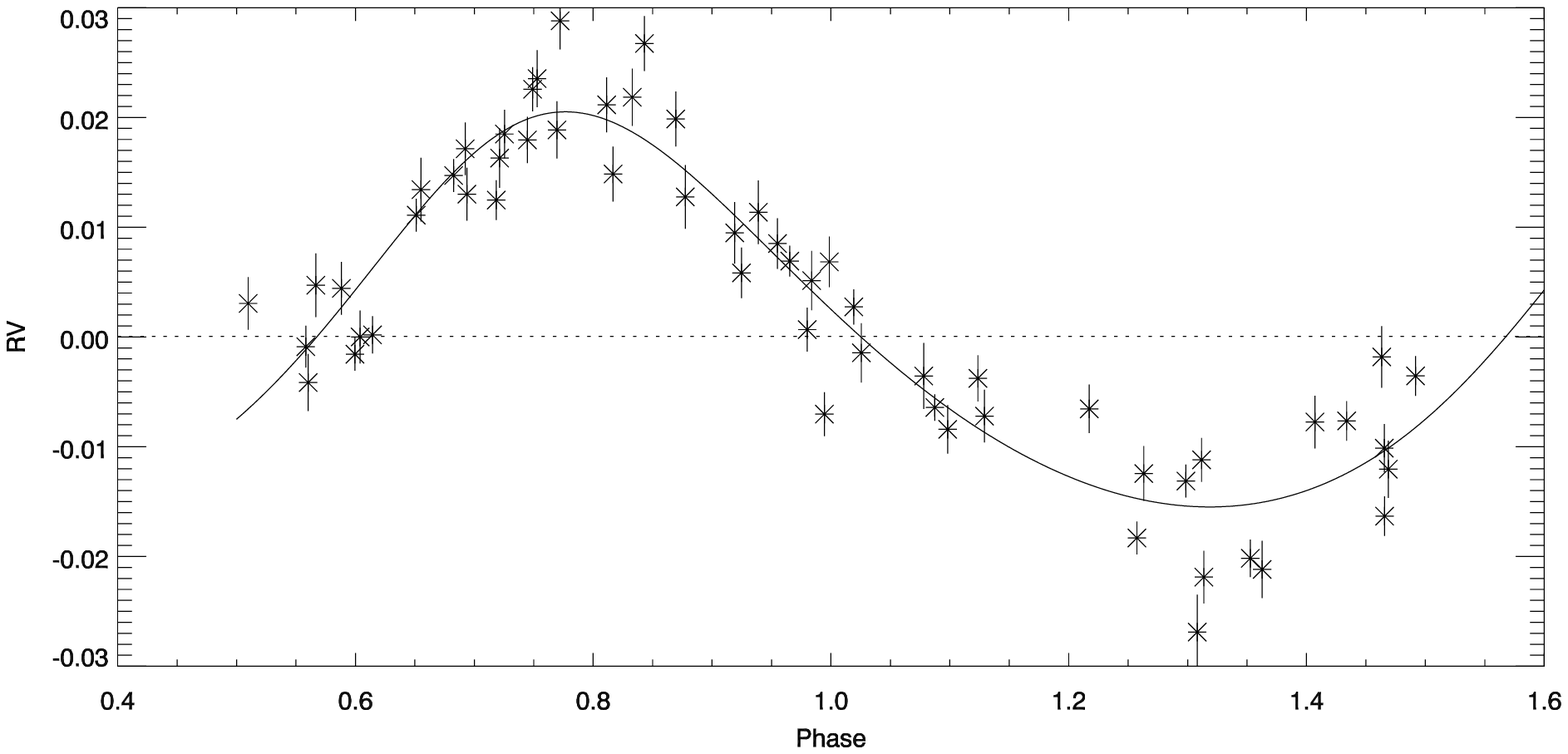,width=\columnwidth,clip=} \\
\caption{Radial velocity data from \citet{man07}, superimposed with the model fit generated by our MCMC analysis. The RV units are in $km\ s^{-1}$. For details, see \S \ref{sec:mcmc}.}
\label{fig:rv}
\end{figure}

We use the following data sets for our analysis:


\begin{itemize}

\item $V$- and $R$-band transit photometry from \citet{gil07b,gil07a}. See Fig. \ref{fig:lcs_1}.
\item {\it Hubble Space Telescope} NICMOS transit photometry from \citet{pon09}. See Fig. \ref{fig:lcs_1}.
\item {\it Spitzer/IRAC4} 8-micron transit and eclipse photometry from \citet{deming07}, \citet{demory07}, and \citet{knu11} as binned by the authors. We adopt the \citet{knu11} parameters used for the correction of their instrumental effects. See Figs. \ref{fig:lcs_1} and \ref{fig:lcs_2}
\item Re-reduced and consolidated RV data presented in \citet{man07} based on data from \citet{but04,but06,man07}, with Julian Data (JD) time stamps converted to heliocentric JD to match the photometry data. See Fig. \ref{fig:rv}.

\end{itemize}

Three publicly available data sets were not included in our analysis due the reasons mentioned below. The higher photometric RMS in the light curves in \citet{cou08} gave this particular data set extremely little weight in the MCMC analysis. The {\it Hubble Space Telescope} F583W transit photometry data from \citet{bea08} are a combination of small 90 minute orbit segments and feature very little out-of-transit data, making it difficult to make appropriate systematic corrections to these segmented data.  The $K$-band transit photometry data from \citet{cac09} have red noise at the level of 1.57 mmag and are missing the pre-transit data, including the first contact point. We note that removing these light curves did not affect the derived properties, but reduced the uncertainties.  In general, the parameters changed by $< 1 \sigma$, except for the impact parameter, which varied by $\sim 2 \sigma$. 





We characterize the system using 10 {\it proposed} parameters that form an approximately orthogonal basis set of uncorrelated parameters that fully characterize the system. Initial guesses are assigned to their values and associated uncertainties. Our set of proposed parameters are the time of minimum light, $T_0$; the orbital period, $P$; the depth of the transit, $\delta$; the duration from first to fourth contact points, $t_T$; the impact parameter, $b$; the values of the eccentricity multiplied by the sine and cosine of the argument of periastron, $e \sin\omega$ and $e \cos\omega$; the semi-amplitude of the RV curve, $K1$; the flux decrement during the secondary eclipse, $\Delta f2$; and our stellar radius, $R_*$.  The proposed stellar radius values at each step are taken from a Gaussian distribution with a mean and $\sigma$ given by our measured value and uncertainty (Table~\ref{tab:properties}), respectively.




At each step in the Markov chain, values for all of the proposed parameters are used to generate model light curves \citep{man02} and radial velocity curves.  Parameter sets are either accepted or rejected based on the $\chi^2$ value when comparing these model curves to the observed time series data, and the accepted parameter values map out the joint posterior probability distribution. The proposed parameters are used to analytically derive the physical parameters for the system, like planet mass and radius, mean stellar density, and stellar mass.
We run the routine for five independent MCMC chains.    Each chain has a 2,000 step burn-in phase and completes after 10,000 accepted steps with an acceptance rate of $\sim 5$\%.   Therefore, we have tested more than a million trial parameter sets to derive the resulting best-fitting parameters and their 1$\sigma$ uncertainties.


\begin{deluxetable}{lcccc}
\tabletypesize{\tiny}
\tablewidth{0pt}
\tablecaption{MCMC Limb Darkening Coefficients\label{tab:coeffs}}
\tablehead{
\colhead{Band} &
\colhead{a1} &
\colhead{a2} &
\colhead{a3} &
\colhead{a4}
}

\startdata

IRAC4 &  0.72352 & -1.01134 & 0.79690 & -0.24584 \\
$V$ &           0.31310 & 1.16686 & -0.85892 & 0.22686 \\
$R$ &           0.37984 & 0.74086 & -0.33350 & 0.018000 \\
$HST$/NICMOS & 	1.533 & -2.234 & 1.913 & -0.643 \\

\enddata

\tablecomments{Limb darkening coefficients used in the MCMC analysis of GJ~436, based on \citet{cla11} for IRAC4, $V$, and $R$, and from \citet{pon09} for $HST$/NICMOS. See \S \ref{sec:analysis}.}

\end{deluxetable}


\subsection{Derived Stellar and Planetary Parameters}\label{sec:parameters}


The results of our MCMC analysis with the directly determined parameters and uncertainties (Table~\ref{tab:properties}) as fixed input values are given in Table~\ref{tab:params}. To provide a graphical insight into the quality of our results, we show, in Figures \ref{fig:lcs_1} through \ref{fig:rv}, the transit/eclipse/RV fits generated by our model, superimposed onto the literature light/RV curves described in \S \ref{sec:analysis}. Further, we illustrate pairwise correlations between individual parameters in Figs. \ref{fig:proposed_corr} and \ref{fig:physical_corr}. 

We do not find any significant deviations in the planet parameters from what has been previously presented in the literature, which is expected given the agreement between our measured stellar radius and the calculated ones, as we discuss in \S \ref{sec:radius}. The two principal differences, however, between our approach and the  ones listed in \S \ref{sec:radius} are (1) we incoporate RV data into our MCMC analysis, and (2) our mass value of GJ~436 ($0.507^{+ 0.071 }_{- 0.062 }$~M$_{\odot}$; Table \ref{tab:params}) is calculated rather than assumed and thus independent of metallicity and other model assumptions. 

\begin{figure}
  \centering
  \includegraphics[angle=0,width=\columnwidth]{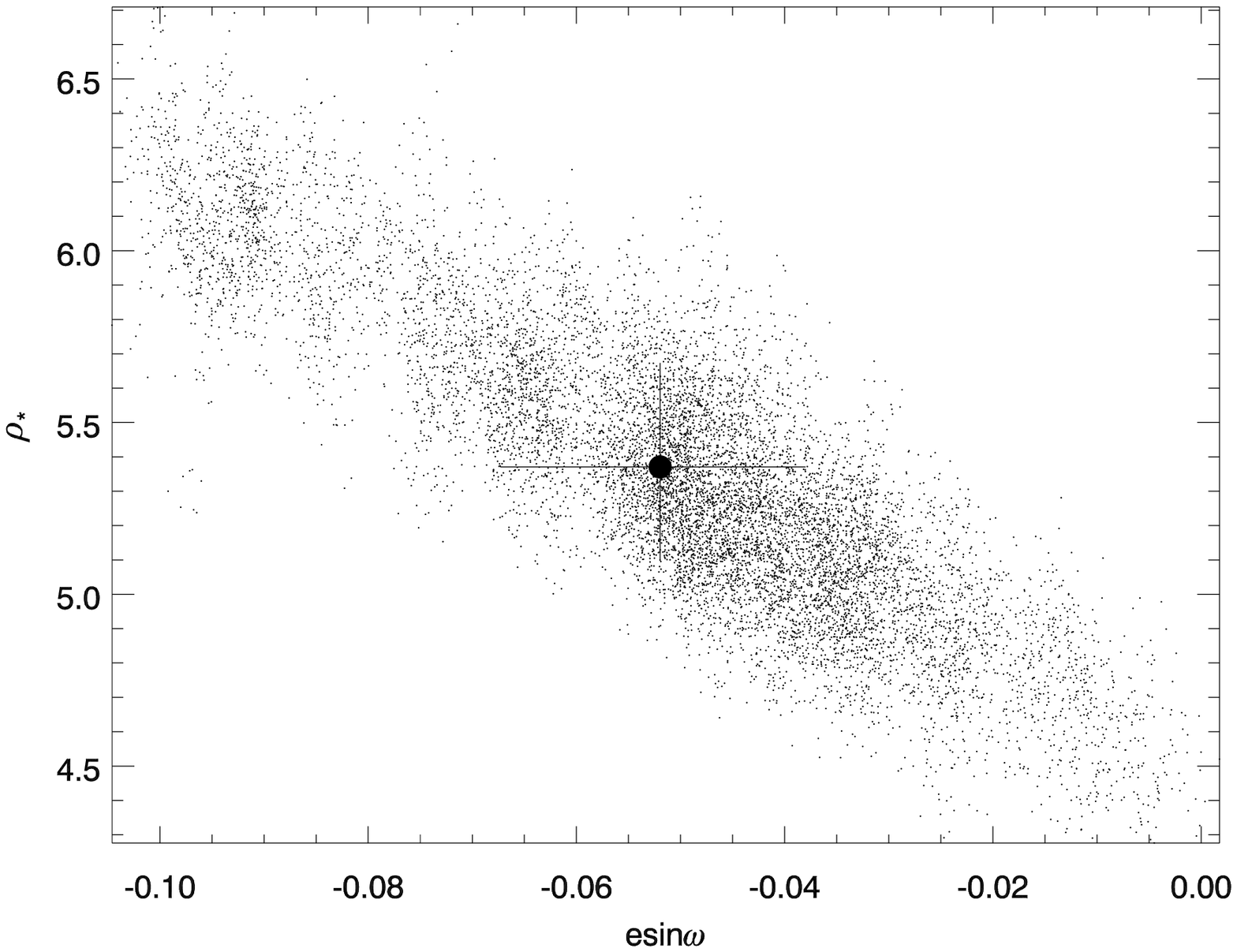}
  \caption{Correlation diagram between $e \sin \omega$ and $\rho_*$ (in units of $\rho_{\odot}$) from our MCMC analysis, along with respective value and uncertainty from Table \ref{tab:params}. $e\sin{\omega}$ has a large associated relative uncertainty that propagates to the mean stellar density and thus causes a relatively large uncertainty in the stellar mass. For details, see \S \ref{sec:parameters} and Table \ref{tab:params}.
  }
  \label{fig:corr_esinw_rho}
\end{figure}

\begin{figure*}
  \centering
  \includegraphics[angle=0,width=\linewidth]{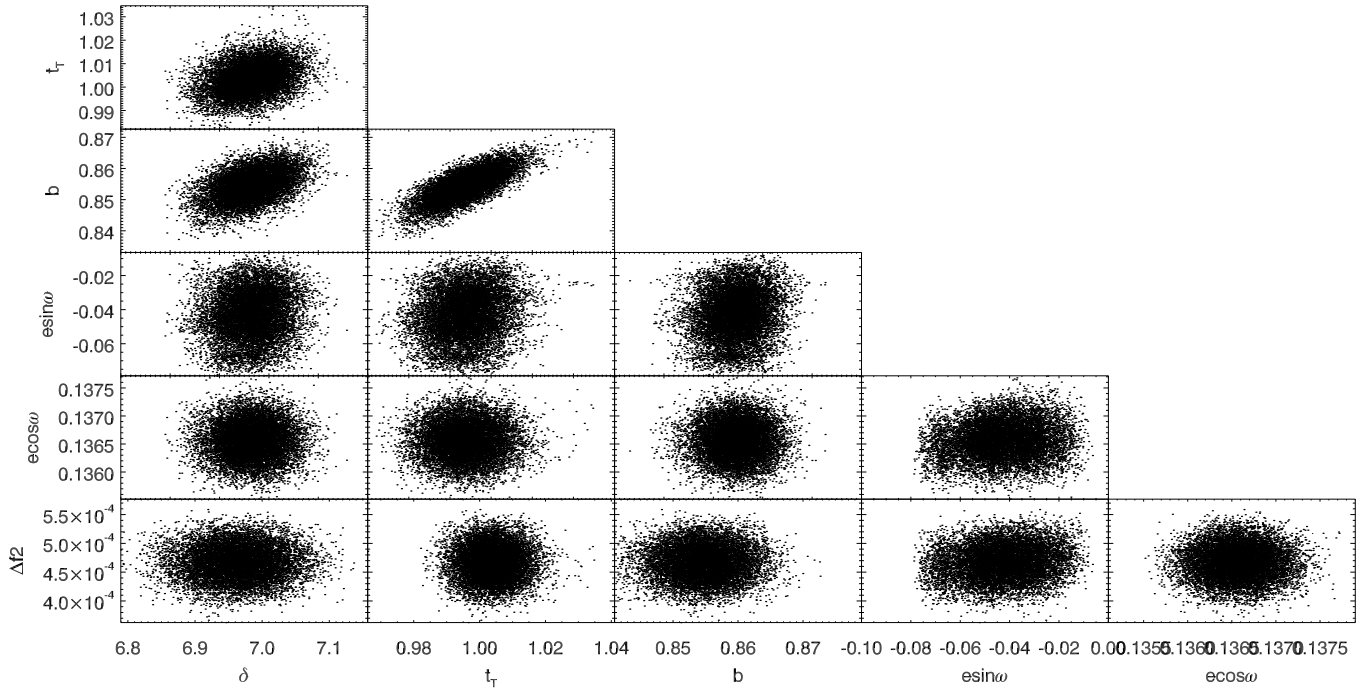}
  \caption{Correlations between some of our proposed parameters of the MCMC analysis. See Table \ref{tab:params} for explanation of symbols and units; $\delta$, the transit depth, is shown here in mmag. For details, see \S \ref{sec:parameters}.
  }
  \label{fig:proposed_corr}
\end{figure*}

\begin{figure*}
  \centering
  \includegraphics[angle=0,width=\linewidth]{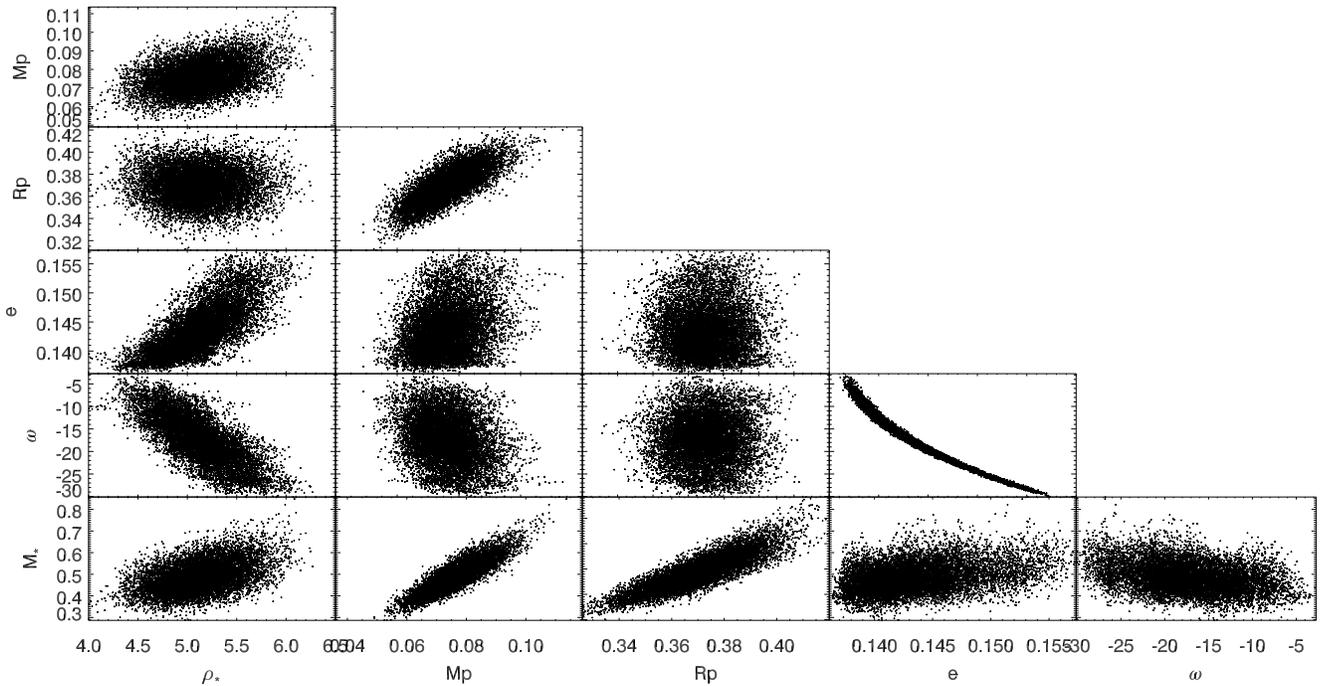}
  \caption{Correlations between some of the GJ~436 physical and orbital system parameters as calculated in our MCMC analysis. See Table \ref{tab:params} for explanation of symbols and units. For details, see \S \ref{sec:parameters}.
  }
  \label{fig:physical_corr}
\end{figure*}

We note that, despite the volume of data used and the superb precision of especially the {\it Spitzer} data, the final (averaged) precision on the stellar mass is only at the level of about 13\%. We believe this is due to the non-zero eccentricity of the system and the high impact parameter, i.e., the almost grazing transit, which both influence the precision of the density measurement. In particular, $e\sin{\omega}$ has a large associated relative uncertainty that propagates through to the stellar mass through its effect on the mean density, as shown in Figure \ref{fig:corr_esinw_rho}.  While the mid-time of the secondary eclipse provides an extremely strong constraint on $e\cos{\omega}$, the degeneracy between the stellar density, $\rho_*$, and $e\sin{\omega}$, means that the secondary eclipse duration provides only a weak constraint on $e\sin{\omega}$.   Furthermore, the radial velocity data is unable to break the degeneracy between eccentricity and $\omega$ due to the small radial velocity amplitude relative to its uncertainties. In Figures \ref{fig:proposed_corr} and \ref{fig:physical_corr}, we show the joint posterior probability distribution of a subset of the final proposed and physical parameters, respectively.  

Finally, it is worth pointing out that our value for $R_p$, calculated from our interferometrically determined stellar radius and the transit depth in the literature light curves (Figures \ref{fig:lcs_1} and \ref{fig:lcs_2}), could be wavelength dependent due to spots on the stellar surface during transit \citep{bal12} or attenuation of stellar radiation in the planetary atmosphere \citep[e.g., fig 14 in][]{knu11}. Our value for $R_p$ is dominated by the {\it Spitzer} 8-micron data due to the number of data points and high photometric precision. Since the magnitude of the aforementioned effects should be smallest at that wavelength range anyway, we calculate a global $\delta$ (Table \ref{tab:params}) and do not let it vary for different data sets.

\begin{deluxetable*}{lccl}

\tablecaption{Derived GJ~436 system parameters and 1$\sigma$ error limits \label{tab:params}}
\tablewidth{0pc}
\tablehead{
\colhead{Parameter} &
\colhead{Symbol} &
\colhead{Value} &
\colhead{Units}	
}

\startdata
Transit epoch (BJD)\tablenotemark{a}\dotfill     & $ T_0  $ & $ 2454510.80096^{+ 0.00005 }_{- 0.00005 } $   	& days \\
Orbital period\tablenotemark{a}\dotfill          & $ P  $   & $ 2.64389826^{+ 0.00000056 }_{- 0.00000058 } $    & days \\
Transit depth\tablenotemark{a}\dotfill  & $ (R_p/R_*)^2 \equiv \delta $ & $ 0.00694^{+ 0.00003 }_{- 0.00003 } $ 	&  \\
Transit duration\tablenotemark{a,}\tablenotemark{b}\dotfill        & $ t_T $ & $ 0.0416^{+ 0.0002 }_{- 0.0002 } $          	& days \\
Impact parameter\tablenotemark{a}\dotfill        & $ b $ & $ 0.853^{+ 0.003 }_{- 0.003 } $                   	& $R_*$ \\
Secondary eclipse depth\tablenotemark{a}\dotfill	& $ \Delta f2 $ & $ 0.00046^{+ 0.00003 }_{- 0.00002 } $ 		& \\
  &    &    &  \\
Stellar reflex velocity\tablenotemark{a}\dotfill & $ K_1 $ & $ 0.018^{+ 0.001 }_{- 0.001 } $             	& km s$^{-1}$ \\
Orbital semimajor axis\dotfill  & $ a $ & $ 0.030^{+ 0.001 }_{- 0.001 } $      			& AU \\
Orbital inclination\dotfill     & $ i $ & $ 86.6^{+ 0.1 }_{- 0.1 } $                		& degrees \\
Orbital eccentricity\dotfill   & $ e $ & $ 0.146^{+ 0.006 }_{- 0.004 } $             		&  \\
Longitude of periastron\dotfill & $ \omega $ & $ -21^{+ 5 }_{- 5 } $            		& degrees  \\
eccentricity $\times \cos(\omega$)\tablenotemark{a} & $e\cos\omega$ & $  0.13653^{+0.00026}_{-0.00027} $            		&  \\
eccentricity $\times \sin(\omega$)\tablenotemark{a} & $e\sin\omega$ & $ -0.05196^{+0.01396}_{-0.01534} $         		&   \\
  &    &    &  \\
Stellar mass\dotfill & $ M_* $      & $ 0.507^{+ 0.071 }_{- 0.062 } $               		& $M_\odot$ \\
Stellar surface gravity\dotfill     & $ \log g_* $ & $ 4.83^{+ 0.03 }_{- 0.03 } $   		& [cgs] \\
Stellar density\dotfill & $ \rho_*$ & $ 5.37^{+ 0.30 }_{- 0.27 } $                   		& $\rho_\odot$ \\
  &    &    &  \\
Planet radius\dotfill & $ R_p $     & $ 0.369^{+ 0.015 }_{- 0.015 } $            			& $R_{Jupiter}$ \\
Planet mass\dotfill & $ M_p $       & $ 0.078^{+ 0.007 }_{- 0.008 } $               			& $M_{Jupiter}$ \\
Planet surface gravity\dotfill   & $ \log g_p $ & $ 3.12^{+ 0.03 }_{- 0.03 } $      	& [cgs] \\
Planet density\dotfill 				& $ \rho_p $ & $ 1.55^{+ 0.12 }_{- 0.10 } $             & $\rho_{Jupiter}$ \\

\enddata

\tablenotetext{a}{Proposed parameters in MCMC analysis (\S \ref{sec:analysis}), along with $R_*$ from Table \ref{tab:properties}.}
\tablenotetext{b}{Defined here as the time between first and fourth contacts.}

\tablecomments{Derived system parameters of GJ~436 from MCMC analysis. Note that the {\it measured} system parameters are given in Table \ref{tab:params}. For details, see \S \ref{sec:mcmc}.}

\end{deluxetable*}


\subsection{Planetary Phase Curve}\label{sec:phasecurve}



In order to ascertain whether GJ~436's thermal phase curve, i.e., the brightness variation as a function of orbital phase of the planet due to the longitudinal surface temperature distribution, could be characterized by observations, we calculate the predicted flux variation of GJ 436 b based upon the parameters shown in Tables \ref{tab:properties} and \ref{tab:params}. We follow the methodology outlined in \citet{kan11} and specifically \citet{kan11a} that takes into account the eccentricity of the orbit. We assume a planet Bond albedo of 0 and no stellar variation for our simulation purposes. 

\begin{figure*}
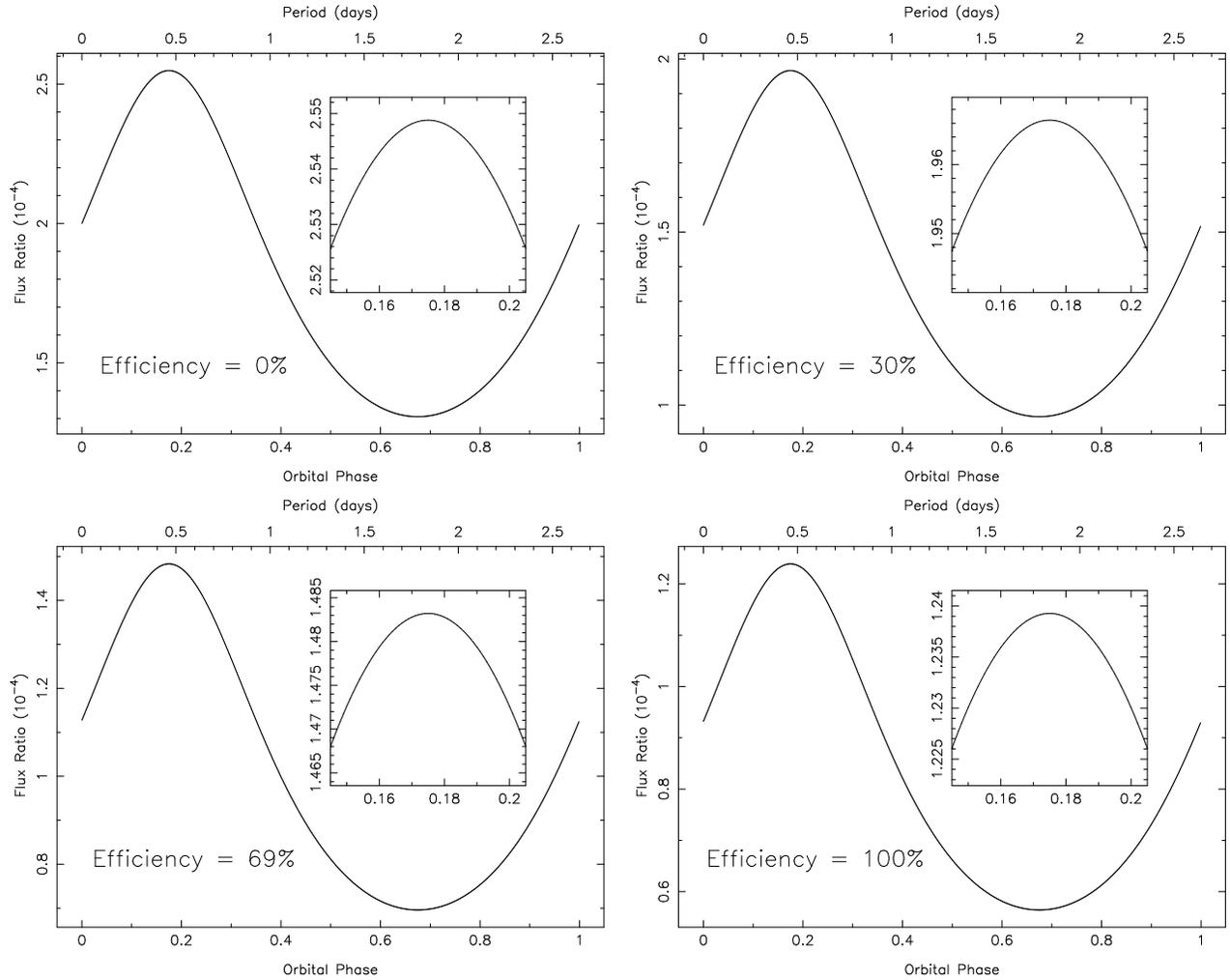

 \begin{center}
   \begin{tabular}{cc}
     \includegraphics[angle=270,width=8.2cm]{phase_0.0.ps} &
     \includegraphics[angle=270,width=8.2cm]{phase_0.3.ps} \\
     \includegraphics[angle=270,width=8.2cm]{phase_0.7.ps} &
     \includegraphics[angle=270,width=8.2cm]{phase_1.0.ps} \\
   \end{tabular}
 \end{center}
 \caption{Simulated phase curve for GJ~436 at 4.5 microns, assuming
 0\%, 30\%, 69\%, and 100\% heat redistribution efficiency of the
 atmosphere \citep[equivalent to the $\eta$ parameter in][]{kan11a}. In each case the sub-panel zooms in on the section of the
 phase curve around the peak in the planet-to-star flux ratio that
 would be optimal for monitoring. For details, see \S \ref{sec:phasecurve}.}
\label{fig:phasecurve}
\vspace{.5 cm}
\end{figure*}

We perform the calculations for variable heat redistribution efficiencies \citep[corresponding to $\eta$ in][]{kan11a} within the planetary atmosphere between 0\% and 100\%. The resulting calculated flux ratio
variations for the various models are shown in Figure \ref{fig:phasecurve}. 

In the case of 100\% heat distribution efficiency (bottom right panel in Fig. \ref{fig:phasecurve}), the phase variation is only
dependent on the changing star-planet distance due to the eccentric orbit, and the amplitude of the phase curve can be regarded as a
lower limit. The eccentricity of the orbit results in a star-planet separation of 0.024 AU at periastron and 0.033 AU at apastron. For 100\% heat redistribution efficiency, we calculate an equilibrium temperature of the planet of 718 K and 611 K at these locations in the orbit, respectively, following the formalism of \citet{skl07}. At a phase angle of zero, where secondary eclipse occurs, we calculate a star-planet separation of 0.027 AU and planetary equilibrium temperature of 675 K. 
The difference between the calculated equilibrium temperature and a value derived from observations during secondary eclipse for the planet day side will be indicative of the magnitude of brightness fluctuations on the planetary surface, potentially caused by the inefficiency of the heat redistribution. From their secondary eclipse observations, \citet{deming07} determine GJ~436's dayside temperature to be 712 K, confirming that any observed phase curve would display larger amplitudes than our simulated one for perfect heat redistribution efficiency, making {\it Spitzer} follow-up observations of this target justified to place constraints on GJ~436's phase curve.



\section{Summary and Conclusion}    \label{sec:conclusion}


In this paper, we present a CHARA-Array interferometric radius for the transiting exoplanet, late-type host star GJ~436. We furthermore calculate a stellar effective temperature based solely on direct measurements. We present the values for these measurements in Table \ref{tab:properties}. We confirm the discrepancy between stellar radii based on stellar models versus direct measurements \citep{boy10} in the M dwarf regime (Boyajian et al. 2012, in preparation), which can also be seen in rapidly rotating, short period eclipsing binaries (EBs) and active late type single stars \citep{yygem2002,mercedes2007,haw96}, as well as inactive field M~dwarfs and long period EBs \citep{boy10,irwin2011}. Calculation of stellar radii based in part on the analysis of transit photometry, however, produces values in good agreement with our interferometric one. 

We use our measured stellar properties in combination with literature time-series data available at the NASA Exoplanet Archive to perform a global analysis of the GJ~436 stellar and planetary system parameters, including radius and mass of the transiting hot Neptune. These calculated parameters are given in Table \ref{tab:params}. Due to the aforementioned agreement between our measured radii and the ones obtained from transit photometry analysis, our calculated system parameters generally agree with literature values.

Planetary characterization is playing an increasingly dominant role in exoplanet research, especially of the nearby and/or bright stellar systems. The parent star obviously dominates the system as the principal energy source, and the object whose interaction with the exoplanet is often all that can be observed to characterize the planet as its own entity. Physical parameters of the planets are thus always a function of their stellar counterparts -- the importance of ``understanding the parent stars'' cannot be overstated. With ongoing improvements in both sensitivity and spatial resolution of near-infrared and optical interferometric data quality, we are able to provide firm, direct measurements of stellar radii and effective temperatures in the low-mass regime to provide a means of comparison to stellar parameters based on transit analysis.


\acknowledgments

We extend our gratitude to the anonymous referee for very insightful remarks on this paper that improved its quality. We furthermore thank J.-P. Beaulieu, D. Kipping, R. Therien, and S. Mahadevan for useful discussions about GJ~436's astrophysical parameters, and especially F. Pont for sending us {\it HST} data from his and collaborators' 2009 study. TSB acknowledges support provided by NASA through Hubble Fellowship grant \#HST-HF-51252.01 awarded by the Space Telescope Science Institute, which is operated by the Association of Universities for Research in Astronomy, Inc., for NASA, under contract NAS 5-26555.  The CHARA Array is funded by the National Science Foundation through NSF grants AST-0606958 and AST-0908253 and by Georgia State University through the College of Arts and Sciences, as well as the W. M. Keck Foundation. This research made use of the SIMBAD literature database, operated at CDS, Strasbourg, France, and of NASA's Astrophysics Data System. This publication makes use of data products from the Two Micron All Sky Survey, which is a joint project of the University of Massachusetts and the Infrared Processing and Analysis Center/California Institute of Technology, funded by the National Aeronautics and Space Administration and the National Science Foundation. This research made use of the NASA Exoplanet Archive, which is operated by the California Institute of Technology, under contract with the National Aeronautics and Space Administration under the Exoplanet Exploration Program.

\newpage




\bibliographystyle{apj}            

\bibliography{apj-jour,paper}      

\begin{thebibliography}{75}
\expandafter\ifx\csname natexlab\endcsname\relax\def\natexlab#1{#1}\fi

\bibitem[{{Baines} {et~al.}(2009){Baines}, {McAlister}, {ten Brummelaar},
  {Sturmann}, {Sturmann}, {Turner}, \& {Ridgway}}]{bai09}
{Baines}, E.~K., {McAlister}, H.~A., {ten Brummelaar}, T.~A., {Sturmann}, J.,
  {Sturmann}, L., {Turner}, N.~H., \& {Ridgway}, S.~T. 2009, \apj, 701, 154

\bibitem[{{Baines} {et~al.}(2010){Baines}, {McAlister}, {ten Brummelaar},
  {Turner}, {Sturmann}, {Sturmann}, {Goldfinger}, {Farrington}, \&
  {Ridgway}}]{bai10}
{Baines}, E.~K. {et~al.} 2010, \aj, 140, 167

\bibitem[{{Baines} {et~al.}(2008{\natexlab{a}}){Baines}, {McAlister}, {ten
  Brummelaar}, {Turner}, {Sturmann}, {Sturmann}, {Goldfinger}, \&
  {Ridgway}}]{bai08}
{Baines}, E.~K., {McAlister}, H.~A., {ten Brummelaar}, T.~A., {Turner}, N.~H.,
  {Sturmann}, J., {Sturmann}, L., {Goldfinger}, P.~J., \& {Ridgway}, S.~T.
  2008{\natexlab{a}}, \apj, 680, 728

\bibitem[{{Baines} {et~al.}(2008{\natexlab{b}}){Baines}, {McAlister}, {ten
  Brummelaar}, {Turner}, {Sturmann}, {Sturmann}, \& {Ridgway}}]{bai08a}
{Baines}, E.~K., {McAlister}, H.~A., {ten Brummelaar}, T.~A., {Turner}, N.~H.,
  {Sturmann}, J., {Sturmann}, L., \& {Ridgway}, S.~T. 2008{\natexlab{b}}, \apj,
  682, 577

\bibitem[{{Baines} {et~al.}(2007){Baines}, {van Belle}, {ten Brummelaar},
  {McAlister}, {Swain}, {Turner}, {Sturmann}, \& {Sturmann}}]{bai07}
{Baines}, E.~K., {van Belle}, G.~T., {ten Brummelaar}, T.~A., {McAlister},
  H.~A., {Swain}, M., {Turner}, N.~H., {Sturmann}, L., \& {Sturmann}, J. 2007,
  \apjl, 661, L195

\bibitem[{{Ballard} {et~al.}(2010{\natexlab{a}}){Ballard}, {Charbonneau},
  {Deming}, {Knutson}, {Christiansen}, {Holman}, {Fabrycky}, {Seager}, \&
  {A'Hearn}}]{bal10b}
{Ballard}, S. {et~al.} 2010{\natexlab{a}}, \pasp, 122, 1341

\bibitem[{{Ballard} {et~al.}(2010{\natexlab{b}}){Ballard}, {Christiansen},
  {Charbonneau}, {Deming}, {Holman}, {Fabrycky}, {A'Hearn}, {Wellnitz},
  {Barry}, {Kuchner}, {Livengood}, {Hewagama}, {Sunshine}, {Hampton}, {Lisse},
  {Seager}, \& {Veverka}}]{bal10a}
------. 2010{\natexlab{b}}, \apj, 716, 1047

\bibitem[{{Ballerini} {et~al.}(2012){Ballerini}, {Micela}, {Lanza}, \&
  {Pagano}}]{bal12}
{Ballerini}, P., {Micela}, G., {Lanza}, A.~F., \& {Pagano}, I. 2012, \aap, 539,
  A140

\bibitem[{{Baraffe} {et~al.}(1998){Baraffe}, {Chabrier}, {Allard}, \&
  {Hauschildt}}]{bar98}
{Baraffe}, I., {Chabrier}, G., {Allard}, F., \& {Hauschildt}, P.~H. 1998, \aap,
  337, 403

\bibitem[{{Bean} {et~al.}(2008){Bean}, {Benedict}, {Charbonneau}, {Homeier},
  {Taylor}, {McArthur}, {Seifahrt}, {Dreizler}, \& {Reiners}}]{bea08}
{Bean}, J.~L. {et~al.} 2008, \aap, 486, 1039

\bibitem[{{Beaulieu} {et~al.}(2011){Beaulieu}, {Tinetti}, {Kipping}, {Ribas},
  {Barber}, {Cho}, {Polichtchouk}, {Tennyson}, {Yurchenko}, {Griffith},
  {Batista}, {Waldmann}, {Miller}, {Carey}, {Mousis}, {Fossey}, \&
  {Aylward}}]{bea11}
{Beaulieu}, J. {et~al.} 2011, \apj, 731, 16

\bibitem[{{Bessell}(2000)}]{bes00}
{Bessell}, M.~S. 2000, \pasp, 112, 961

\bibitem[{{Boyajian} {et~al.}(2011){Boyajian}, {McAlister}, {van Belle},
  {Gies}, {ten Brummelaar}, {von Braun}, {Farrington}, {Goldfinger}, {O'Brien},
  {Parks}, {Richardson}, {Ridgway}, {Schaefer}, {Sturmann}, {Sturmann},
  {Touhami}, {Turner}, \& {White}}]{boy11}
{Boyajian}, T.~S. {et~al.} 2011, ArXiv e-prints; astro-ph/1112.3316

\bibitem[{{Boyajian} {et~al.}(2010){Boyajian}, {von Braun}, {van Belle}, {ten
  Brummelaar}, {Ciardi}, {Henry}, {Lopez-Morales}, {McAlister}, {Ridgway},
  {Farrington}, {Goldfinger}, {Sturmann}, {Sturmann}, \& {Turner}}]{boy10}
------. 2010, astro-ph/1012.0542

\bibitem[{{Butler} {et~al.}(2004){Butler}, {Vogt}, {Marcy}, {Fischer},
  {Wright}, {Henry}, {Laughlin}, \& {Lissauer}}]{but04}
{Butler}, R.~P., {Vogt}, S.~S., {Marcy}, G.~W., {Fischer}, D.~A., {Wright},
  J.~T., {Henry}, G.~W., {Laughlin}, G., \& {Lissauer}, J.~J. 2004, \apj, 617,
  580

\bibitem[{{Butler} {et~al.}(2006){Butler}, {Wright}, {Marcy}, {Fischer},
  {Vogt}, {Tinney}, {Jones}, {Carter}, {Johnson}, {McCarthy}, \&
  {Penny}}]{but06}
{Butler}, R.~P. {et~al.} 2006, \apj, 646, 505

\bibitem[{{C{\'a}ceres} {et~al.}(2009){C{\'a}ceres}, {Ivanov}, {Minniti},
  {Naef}, {Melo}, {Mason}, {Selman}, \& {Pietrzynski}}]{cac09}
{C{\'a}ceres}, C., {Ivanov}, V.~D., {Minniti}, D., {Naef}, D., {Melo}, C.,
  {Mason}, E., {Selman}, F., \& {Pietrzynski}, G. 2009, \aap, 507, 481

\bibitem[{{Claret}(2000)}]{cla00}
{Claret}, A. 2000, \aap, 363, 1081

\bibitem[{{Claret} \& {Bloemen}(2011)}]{cla11}
{Claret}, A., \& {Bloemen}, S. 2011, \aap, 529, A75

\bibitem[{{Collier Cameron} {et~al.}(2007){Collier Cameron}, {Wilson}, {West},
  {Hebb}, {Wang}, {Aigrain}, {Bouchy}, {Christian}, {Clarkson}, {Enoch},
  {Esposito}, {Guenther}, {Haswell}, {H{\'e}brard}, {Hellier}, {Horne},
  {Irwin}, {Kane}, {Loeillet}, {Lister}, {Maxted}, {Mayor}, {Moutou}, {Parley},
  {Pollacco}, {Pont}, {Queloz}, {Ryans}, {Skillen}, {Street}, {Udry}, \&
  {Wheatley}}]{cameron2007}
{Collier Cameron}, A. {et~al.} 2007, \mnras, 380, 1230

\bibitem[{{Coughlin} {et~al.}(2008){Coughlin}, {Stringfellow}, {Becker},
  {L{\'o}pez-Morales}, {Mezzalira}, \& {Krajci}}]{cou08}
{Coughlin}, J.~L., {Stringfellow}, G.~S., {Becker}, A.~C., {L{\'o}pez-Morales},
  M., {Mezzalira}, F., \& {Krajci}, T. 2008, \apjl, 689, L149

\bibitem[{{Cutri} {et~al.}(2003){Cutri}, {Skrutskie}, {van Dyk}, {Beichman},
  {Carpenter}, {Chester}, {Cambresy}, {Evans}, {Fowler}, {Gizis}, {Howard},
  {Huchra}, {Jarrett}, {Kopan}, {Kirkpatrick}, {Light}, {Marsh}, {McCallon},
  {Schneider}, {Stiening}, {Sykes}, {Weinberg}, {Wheaton}, {Wheelock}, \&
  {Zacarias}}]{cut03}
{Cutri}, R.~M. {et~al.} 2003, {The 2MASS All Sky Catalog of Point Sources}
  (Pasadena: IPAC)

\bibitem[{{Delfosse} {et~al.}(2000){Delfosse}, {Forveille}, {S{\'e}gransan},
  {Beuzit}, {Udry}, {Perrier}, \& {Mayor}}]{del00}
{Delfosse}, X., {Forveille}, T., {S{\'e}gransan}, D., {Beuzit}, J., {Udry}, S.,
  {Perrier}, C., \& {Mayor}, M. 2000, \aap, 364, 217

\bibitem[{{Deming} {et~al.}(2007){Deming}, {Harrington}, {Laughlin}, {Seager},
  {Navarro}, {Bowman}, \& {Horning}}]{deming07}
{Deming}, D., {Harrington}, J., {Laughlin}, G., {Seager}, S., {Navarro}, S.~B.,
  {Bowman}, W.~C., \& {Horning}, K. 2007, \apjl, 667, L199

\bibitem[{{Demory} {et~al.}(2007){Demory}, {Gillon}, {Barman}, {Bonfils},
  {Mayor}, {Mazeh}, {Queloz}, {Udry}, {Bouchy}, {Delfosse}, {Forveille},
  {Mallmann}, {Pepe}, \& {Perrier}}]{demory07}
{Demory}, B. {et~al.} 2007, \aap, 475, 1125

\bibitem[{{Demory} {et~al.}(2009){Demory}, {S{\'e}gransan}, {Forveille},
  {Queloz}, {Beuzit}, {Delfosse}, {di Folco}, {Kervella}, {Le Bouquin},
  {Perrier}, {Benisty}, {Duvert}, {Hofmann}, {Lopez}, \& {Petrov}}]{dem09}
{Demory}, B.-O. {et~al.} 2009, \aap, 505, 205

\bibitem[{{Doinidis} \& {Beers}(1991)}]{doi91}
{Doinidis}, S.~P., \& {Beers}, T.~C. 1991, \pasp, 103, 973

\bibitem[{{Enoch} {et~al.}(2010){Enoch}, {Collier Cameron}, {Parley}, \&
  {Hebb}}]{eno10}
{Enoch}, B., {Collier Cameron}, A., {Parley}, N.~R., \& {Hebb}, L. 2010, \aap,
  516, A33+

\bibitem[{{Figueira} {et~al.}(2009){Figueira}, {Pont}, {Mordasini}, {Alibert},
  {Georgy}, \& {Benz}}]{fig09}
{Figueira}, P., {Pont}, F., {Mordasini}, C., {Alibert}, Y., {Georgy}, C., \&
  {Benz}, W. 2009, \aap, 493, 671

\bibitem[{{Gillon} {et~al.}(2007{\natexlab{a}}){Gillon}, {Demory}, {Barman},
  {Bonfils}, {Mazeh}, {Pont}, {Udry}, {Mayor}, \& {Queloz}}]{gil07b}
{Gillon}, M. {et~al.} 2007{\natexlab{a}}, \aap, 471, L51

\bibitem[{{Gillon} {et~al.}(2007{\natexlab{b}}){Gillon}, {Pont}, {Demory},
  {Mallmann}, {Mayor}, {Mazeh}, {Queloz}, {Shporer}, {Udry}, \&
  {Vuissoz}}]{gil07a}
------. 2007{\natexlab{b}}, \aap, 472, L13

\bibitem[{{Golay}(1972)}]{gol72}
{Golay}, M. 1972, Vistas in Astronomy, 14, 13

\bibitem[{{Hanbury Brown} {et~al.}(1974){Hanbury Brown}, {Davis}, {Lake}, \&
  {Thompson}}]{han74}
{Hanbury Brown}, R., {Davis}, J., {Lake}, R.~J.~W., \& {Thompson}, R.~J. 1974,
  \mnras, 167, 475

\bibitem[{{Hawley} {et~al.}(1996){Hawley}, {Gizis}, \& {Reid}}]{haw96}
{Hawley}, S.~L., {Gizis}, J.~E., \& {Reid}, I.~N. 1996, \aj, 112, 2799

\bibitem[{{Hebb} {et~al.}(2010){Hebb}, {Collier-Cameron}, {Triaud}, {Lister},
  {Smalley}, {Maxted}, {Hellier}, {Anderson}, {Pollacco}, {Gillon}, {Queloz},
  {West}, {Bentley}, {Enoch}, {Haswell}, {Horne}, {Mayor}, {Pepe}, {Segransan},
  {Skillen}, {Udry}, \& {Wheatley}}]{heb10}
{Hebb}, L. {et~al.} 2010, \apj, 708, 224

\bibitem[{{Irwin} {et~al.}(2011){Irwin}, {Quinn}, {Berta}, {Latham}, {Torres},
  {Burke}, {Charbonneau}, {Dittmann}, {Esquerdo}, {Stefanik}, {Oksanen},
  {Buchhave}, {Nutzman}, {Berlind}, {Calkins}, \& {Falco}}]{irwin2011}
{Irwin}, J.~M. {et~al.} 2011, ArXiv e-prints

\bibitem[{{Kane} {et~al.}(2011){Kane}, {Ciardi}, {Dragomir}, {Gelino}, \& {von
  Braun}}]{kan11}
{Kane}, S.~R., {Ciardi}, D.~R., {Dragomir}, D., {Gelino}, D.~M., \& {von
  Braun}, K. 2011, ArXiv e-prints; astro-ph/1105.1716

\bibitem[{{Kane} \& {Gelino}(2011)}]{kan11a}
{Kane}, S.~R., \& {Gelino}, D.~M. 2011, ArXiv e-prints

\bibitem[{{Kirkpatrick} {et~al.}(1991){Kirkpatrick}, {Henry}, \&
  {McCarthy}}]{kir91}
{Kirkpatrick}, J.~D., {Henry}, T.~J., \& {McCarthy}, Jr., D.~W. 1991, \apjs,
  77, 417

\bibitem[{{Knutson} {et~al.}(2011){Knutson}, {Madhusudhan}, {Cowan},
  {Christiansen}, {Agol}, {Deming}, {D{\'e}sert}, {Charbonneau}, {Henry},
  {Homeier}, {Langton}, {Laughlin}, \& {Seager}}]{knu11}
{Knutson}, H.~A. {et~al.} 2011, \apj, 735, 27

\bibitem[{{Leggett}(1992)}]{leg92}
{Leggett}, S.~K. 1992, \apjs, 82, 351

\bibitem[{{L{\'o}pez-Morales}(2007)}]{lop07}
{L{\'o}pez-Morales}, M. 2007, \apj, 660, 732

\bibitem[{{L{\'o}pez-Morales} \& {Shaw}(2007)}]{mercedes2007}
{L{\'o}pez-Morales}, M., \& {Shaw}, J.~S. 2007, in Astronomical Society of the
  Pacific Conference Series, Vol. 362, The Seventh Pacific Rim Conference on
  Stellar Astrophysics, ed. {Y.~W.~Kang, H.-W.~Lee, K.-C.~Leung, \&
  K.-S.~Cheng}, 26--+

\bibitem[{{Mandel} \& {Agol}(2002)}]{man02}
{Mandel}, K., \& {Agol}, E. 2002, \apjl, 580, L171

\bibitem[{{Maness} {et~al.}(2007){Maness}, {Marcy}, {Ford}, {Hauschildt},
  {Shreve}, {Basri}, {Butler}, \& {Vogt}}]{man07}
{Maness}, H.~L., {Marcy}, G.~W., {Ford}, E.~B., {Hauschildt}, P.~H., {Shreve},
  A.~T., {Basri}, G.~B., {Butler}, R.~P., \& {Vogt}, S.~S. 2007, \pasp, 119, 90

\bibitem[{{Mermilliod}(1986)}]{mer86}
{Mermilliod}, J.-C. 1986, Catalogue of Eggen's UBV data., 0 (1986), 0

\bibitem[{{Pickles}(1998)}]{pic98}
{Pickles}, A.~J. 1998, \pasp, 110, 863

\bibitem[{{Pont} {et~al.}(2009){Pont}, {Gilliland}, {Knutson}, {Holman}, \&
  {Charbonneau}}]{pon09}
{Pont}, F., {Gilliland}, R.~L., {Knutson}, H., {Holman}, M., \& {Charbonneau},
  D. 2009, \mnras, 393, L6

\bibitem[{{Rojas-Ayala} {et~al.}(2010){Rojas-Ayala}, {Covey}, {Muirhead}, \&
  {Lloyd}}]{roj10}
{Rojas-Ayala}, B., {Covey}, K.~R., {Muirhead}, P.~S., \& {Lloyd}, J.~P. 2010,
  \apjl, 720, L113

\bibitem[{{Rufener}(1976)}]{ruf76}
{Rufener}, F. 1976, \aaps, 26, 275

\bibitem[{{Seager} \& {Mall{\'e}n-Ornelas}(2003)}]{sea03}
{Seager}, S., \& {Mall{\'e}n-Ornelas}, G. 2003, \apj, 585, 1038

\bibitem[{{Selsis} {et~al.}(2007){Selsis}, {Kasting}, {Levrard}, {Paillet},
  {Ribas}, \& {Delfosse}}]{skl07}
{Selsis}, F., {Kasting}, J.~F., {Levrard}, B., {Paillet}, J., {Ribas}, I., \&
  {Delfosse}, X. 2007, \aap, 476, 1373

\bibitem[{{Shporer} {et~al.}(2009){Shporer}, {Mazeh}, {Pont}, {Winn}, {Holman},
  {Latham}, \& {Esquerdo}}]{shp09}
{Shporer}, A., {Mazeh}, T., {Pont}, F., {Winn}, J.~N., {Holman}, M.~J.,
  {Latham}, D.~W., \& {Esquerdo}, G.~A. 2009, \apj, 694, 1559

\bibitem[{{Southworth}(2008)}]{sou08a}
{Southworth}, J. 2008, \mnras, 386, 1644

\bibitem[{{Southworth}(2009)}]{sou08b}
------. 2009, \mnras, 394, 272

\bibitem[{{Southworth}(2010)}]{sou10}
------. 2010, \mnras, 408, 1689

\bibitem[{{Sozzetti} {et~al.}(2007){Sozzetti}, {Torres}, {Charbonneau},
  {Latham}, {Holman}, {Winn}, {Laird}, \& {O'Donovan}}]{soz07}
{Sozzetti}, A., {Torres}, G., {Charbonneau}, D., {Latham}, D.~W., {Holman},
  M.~J., {Winn}, J.~N., {Laird}, J.~B., \& {O'Donovan}, F.~T. 2007, \apj, 664,
  1190

\bibitem[{{Stauffer} \& {Hartmann}(1986)}]{sta86}
{Stauffer}, J.~R., \& {Hartmann}, L.~W. 1986, \apjs, 61, 531

\bibitem[{{Stevenson} {et~al.}(2010){Stevenson}, {Harrington}, {Nymeyer},
  {Madhusudhan}, {Seager}, {Bowman}, {Hardy}, {Deming}, {Rauscher}, \&
  {Lust}}]{ste10}
{Stevenson}, K.~B. {et~al.} 2010, \nat, 464, 1161

\bibitem[{{Sturmann} {et~al.}(2003){Sturmann}, {ten Brummelaar}, {Ridgway},
  {Shure}, {Safizadeh}, {Sturmann}, {Turner}, \& {McAlister}}]{stu03}
{Sturmann}, J., {ten Brummelaar}, T.~A., {Ridgway}, S.~T., {Shure}, M.~A.,
  {Safizadeh}, N., {Sturmann}, L., {Turner}, N.~H., \& {McAlister}, H.~A. 2003,
  in Society of Photo-Optical Instrumentation Engineers (SPIE) Conference
  Series, Vol. 4838, Society of Photo-Optical Instrumentation Engineers (SPIE)
  Conference Series, ed. {W.~A.~Traub}, 1208--1215

\bibitem[{{ten Brummelaar} {et~al.}(2005){ten Brummelaar}, {McAlister},
  {Ridgway}, {Bagnuolo}, {Turner}, {Sturmann}, {Sturmann}, {Berger}, {Ogden},
  {Cadman}, {Hartkopf}, {Hopper}, \& {Shure}}]{ten05}
{ten Brummelaar}, T.~A. {et~al.} 2005, \apj, 628, 453

\bibitem[{{Tingley} {et~al.}(2011){Tingley}, {Bonomo}, \& {Deeg}}]{tin11}
{Tingley}, B., {Bonomo}, A.~S., \& {Deeg}, H.~J. 2011, \apj, 726, 112

\bibitem[{{Torres}(2007)}]{tor07}
{Torres}, G. 2007, \apjl, 671, L65

\bibitem[{{Torres} \& {Ribas}(2002)}]{yygem2002}
{Torres}, G., \& {Ribas}, I. 2002, \apj, 567, 1140

\bibitem[{{van Belle}(2008)}]{van08a}
{van Belle}, G.~T. 2008, \pasp, 120, 617

\bibitem[{{van Belle} \& {von Braun}(2009)}]{van09}
{van Belle}, G.~T., \& {von Braun}, K. 2009, \apj, 694, 1085

\bibitem[{{van Leeuwen}(2007)}]{van07}
{van Leeuwen}, F. 2007, {Hipparcos, the New Reduction of the Raw Data}
  (Hipparcos, the New Reduction of the Raw Data.~By Floor van Leeuwen,
  Institute of Astronomy, Cambridge University, Cambridge, UK Series:
  Astrophysics and Space Science Library, Vol.~ 350 20 Springer Dordrecht)

\bibitem[{{von Braun} {et~al.}(2011{\natexlab{a}}){von Braun}, {Boyajian},
  {Kane}, {van Belle}, {Ciardi}, {L{\'o}pez-Morales}, {McAlister}, {Henry},
  {Jao}, {Riedel}, {Subasavage}, {Schaefer}, {ten Brummelaar}, {Ridgway},
  {Sturmann}, {Sturmann}, {Mazingue}, {Turner}, {Farrington}, {Goldfinger}, \&
  {Boden}}]{von11a}
{von Braun}, K. {et~al.} 2011{\natexlab{a}}, \apjl, 729, L26+

\bibitem[{{von Braun} {et~al.}(2011{\natexlab{b}}){von Braun}, {Boyajian}, {ten
  Brummelaar}, {van Belle}, {Kane}, {Ciardi}, {Lopez-Morales}, {McAlister},
  {Schaefer}, {Ridgway}, {Sturmann}, {Sturmann}, {White}, {Turner},
  {Farrington}, \& {Goldfinger}}]{von11b}
------. 2011{\natexlab{b}}, ArXiv e-prints; astro-ph/1107.1936

\bibitem[{{von Braun} {et~al.}(2011{\natexlab{c}}){von Braun}, {Boyajian
  Tabetha}, {ten Brummelaar}, {Kane}, {van Belle}, {Ciardi}, {Raymond},
  {L{\'o}pez-Morales}, {McAlister}, {Schaefer}, {Ridgway}, {Sturmann},
  {Sturmann}, {White}, {Turner}, {Farrington}, \& {Goldfinger}}]{von11c}
------. 2011{\natexlab{c}}, \apj, 740, 49

\bibitem[{{von Braun} {et~al.}(2008){von Braun}, {van Belle}, {Ciardi},
  {L{\'o}pez-Morales}, {Hoard}, \& {Wachter}}]{von08}
{von Braun}, K., {van Belle}, G.~T., {Ciardi}, D.~R., {L{\'o}pez-Morales}, M.,
  {Hoard}, D.~W., \& {Wachter}, S. 2008, \apj, 677, 545

\bibitem[{{Weis}(1993)}]{wei93}
{Weis}, E.~W. 1993, \aj, 105, 1962

\bibitem[{{Weis}(1996)}]{wei96}
------. 1996, \aj, 112, 2300

\end{thebibliography}





\end{document}